\documentclass[11pt,a4paper]{article}
\usepackage{jcappub}
\usepackage{bm}
\usepackage{epsfig}
\usepackage{graphicx}
\usepackage{amssymb}
\usepackage{xcolor}
\usepackage{subfigure}
\renewcommand\({\left(}
\renewcommand\){\right)}

\newcommand{\be}{\begin{equation}}
\newcommand{\ee}{\end{equation}}
\newcommand{\bea}{\begin{eqnarray}}
\newcommand{\eea}{\end{eqnarray}}

\renewcommand{\deg}{$^{\circ}$ }

\subheader{\hfill MPP-2014-48, DESY-2014-xx}

\title{Results from the Solar Hidden Photon Search (SHIPS)}

\author[a]{Matthias Schwarz} 
\author[b]{Ernst-Axel Knabbe}
\author[b]{Axel Lindner}
\author[c,d]{Javier Redondo}
\author[b]{Andreas Ringwald}
\author[a]{Magnus Schneide}
\author[a]{Jaroslaw Susol}
\author[a]{G\"unter Wiedemann}

\affiliation[a]{Hamburger Sternwarte, Gojenbergsweg 112, 21029 Hamburg, Germany}

\affiliation[b]{Deutsches Elektronen-Synchrotron DESY, Notkestra\ss e 85, 22607 Hamburg, Germany}

\affiliation[c]{Departamento de F\'isica Te\'orica, Universidad de Zaragoza,\\ Pedro Cerbuna 12, E-50009, Zaragoza, Espa\~na.}

\affiliation[d]{Max-Planck-Institut f\"ur Physik
(Werner-Heisenberg-Institut)\\
F\"ohringer Ring 6, D-80805 M\"unchen, Germany}

\abstract{
We present the results of a search for transversely polarised hidden photons (HPs) with $\sim 3$ eV energies emitted from the Sun. 
These hypothetical particles, known also as paraphotons or dark sector photons, are theoretically well motivated for example by string theory inspired extensions of the Standard Model.
Solar HPs of sub-eV mass can convert into photons of the same energy (photon$\leftrightarrow$HP oscillations are similar to neutrino flavour oscillations). 
At SHIPS this would take place inside a long light-tight high-vacuum tube, which tracks the Sun. The generated photons would then be focused into a low-noise photomultiplier at the far end of the tube. Our analysis of 330 h of data (and \mbox{330 h} of background characterisation) reveals no signal of photons from solar hidden photon conversion. We estimate the rate of newly generated photons due to this conversion to be smaller than 25 mHz/m$^2$ at the 95$\%$ C.L. Using this and a recent model of solar HP emission, we set stringent constraints on $\chi$, the coupling constant between HPs and photons, as a function of the HP mass.}


\begin{document}
\maketitle

%\end{abstract}

%%%%%%%%%%%%%%%%%%%%%%%%%%%%%%%%%%%%%%%%%%%%%%%%%%%%%%%%%%%%%%%%%%%%%%
\section{Introduction}                        \label{sec:introduction}
%%%%%%%%%%%%%%%%%%%%%%%%%%%%%%%%%%%%%%%%%%%%%%%%%%%%%%%%%%%%%%%%%%%%%%

In 1982, L. D. Okun proposed a theory to investigate the precision with which we test quantum electrodynamics (QED) in a quest of identifying new fundamental forces of nature~\cite{Okun:1982xi}. The theory consisted of adding an extra vector field, dubbed paraphoton, with a small mass and also small coupling to the electric charge. These parameters can be constrained by precision QED tests because the paraphoton mass modifies Coulomb's $1/r^2$ law and the morphology of magnetic fields proportionally to the new coupling. 
However, it turned out that the biggest impact of this theory was not due to the new macroscopic force, but to the production of the quanta of the new field. 
In this theory, the electric charge excites one particular linear combination of the standard photon and the paraphoton field, and this implies that the orthogonal combination is not excited, nor it can be absorbed: it becomes sterile. Neither this sterile photon nor the combination excited by the charge are pure mass eigenstates, so electromagnetic radiation oscillates into the sterile state and back as it propagates. This mechanism allowed Okun to predict a number of very exotic phenomena such as spectral distortions of astrophysical sources (the oscillation probability is frequency dependent), light-shining-through walls and the emission of a huge flux of paraphotons from the Sun, which he speculated could be the target of very sensitive experiments.  

Having such a low mass and weakly coupled paraphoton was not particularly well motivated theoretically until Holdom and others showed that precisely these couplings arise as a result of kinetic mixing~\cite{Holdom:1985ag,Foot:1991kb,Foot:1991bp} between the photon and the paraphoton fields in the Lagrangian density, 
\be
{\cal L} \ni \frac{1}{2}\chi F_{\mu\nu}X^{\mu\nu},
\ee
where $F_{\mu\nu}=\partial_\mu A_\nu-\partial_\nu A_\mu$ is the photon field strength, $A_\mu$ is the photon field, $X_{\mu\nu}$ is the equivalent for the paraphoton, and $\chi$ is the kinetic mixing parameter. 
Here the paraphoton (also dubbed as hidden photon (HP)) is simply the abelian vector boson of a hidden U(1)$_h$ symmetry, under which the known particles are uncharged. Kinetic mixing is generated by a radiative correction of typical size $\chi\sim O(10^{-4})$, although it can be much smaller if any/both of the involved U(1)'s is embedded into a non-abelian gauge group or if the hidden gauge coupling is tiny. Interestingly, hidden U(1) symmetries appear copiously in completions of the standard model of particle physics based on string theory. The sizes of the HP mass, $m$, and kinetic mixing, $\chi$, have been the subject of intense studies ever since~\cite{Dienes:1996zr,Lukas:1999nh,Abel:2003ue,Blumenhagen:2005ga,Abel:2006qt,Abel:2008ai,Goodsell:2009pi,Goodsell:2009xc,Goodsell:2010ie,Heckman:2010fh,Bullimore:2010aj,Cicoli:2011yh,Goodsell:2011wn}. Predictions span a huge range in parameter space, see~\cite{Jaeckel:2010ni,Jaeckel:2013ija,Ringwald:2012hr} for recent reviews. It is intriguing that 
with finding these low mass particles one can 
learn (at least in some cases) about the central parameters of a more fundamental theory of nature, realised completely only at the highest energies.  

The experimental search of hidden photons \`a la Okun has followed these theoretical advances, gathering recently a lot of attention. 
Spectral distortions of the cosmic microwave background (CMB) caused by photon-hidden photon oscillations were searched in~\cite{Georgi:1983sy,Nordberg:1998wn,Mirizzi:2009iz,Jaeckel:2008fi}, resulting in very strong constraints. Searches in the radio regime were also recently proposed~\cite{Lobanov:2012pt}. 

\begin{figure}[tbp]
\begin{center}
\includegraphics[width=7.7cm]{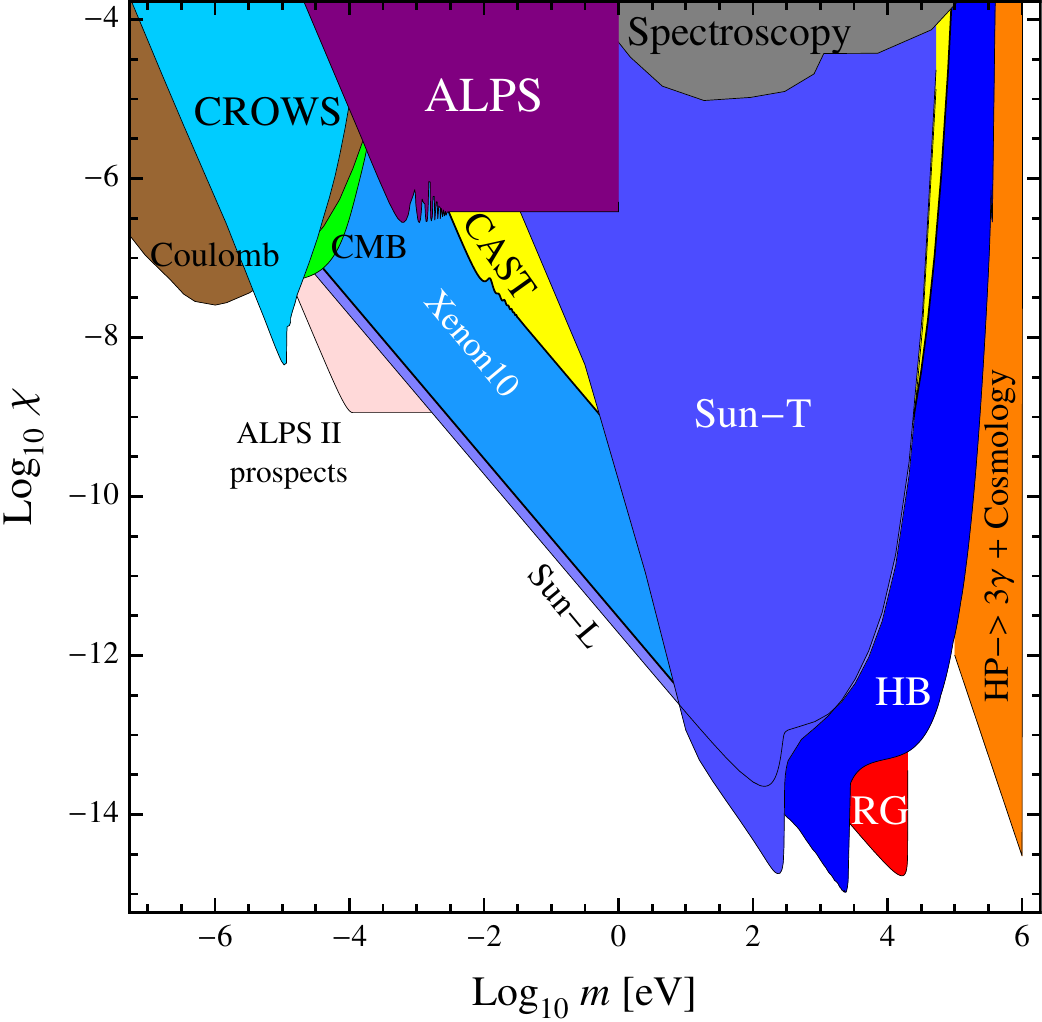}
\vspace{-.3cm}
\caption{\small Experimental constraints on the kinetic mixing of hypothetical HPs with ordinary photons, as a function of the HP mass. See~\cite{Redondo:2013lna,Jaeckel:2013ija,Redondo:2014,Vinyoles:2015aba} for references.}
\label{bounds}
\vspace{-.7cm}
\end{center}
\end{figure}

A number of light-shining-through-walls experiments were also performed. First, with laser light around the visible~\cite{Cameron:1993mr,Fouche:2008jk,Afanasev:2008fv,Ehret:2010mh}, where the Any-Light-Particle-Search (ALPS) at DESY ~\cite{Ehret:2010mh} is currently the most sensitive and already prepares a next generation (ALPS II)~\cite{Bahre:2013ywa}. Later, also with microwaves~\cite{Wagner:2010mi,Betz:2013dza,Parker:2013fxa} as suggested in~\cite{Jaeckel:2007ch}, where the sensitivity at low HP masses (below 0.1 meV) is enormously improved over laser experiments. Here, the CROWS experi\-ment~\cite{Betz:2013dza} is currently much better than ALPS, and recent ideas~\cite{Graham:2014sha} promise even further improvements. 
Tests of the Coulomb $1/r^2$ law at macroscopic distances have not been improved since the 70's~\cite{Bartlett:1970js,Williams:1971ms} but at atomic distances new constraints on HPs were recently derived~\cite{Jaeckel:2010xx}. A summary of experimental constrains on $\chi$ as a function of the HP mass, $m$, is shown in Fig.~\ref{bounds}. 

Searches for solar HPs provide the dominant constraints on the kinetic mixing in the broad region of HP masses between meV and 10 keV.  
Okun himself estimated the constraints for the HP flux not to exceed the standard solar luminosity in photons. 
Later, they were refined by Popov and Vasil'ev~\cite{Popov:1991} and by Popov~\cite{Popov:1999}. 
First experiments looking directly for the solar HP flux were reported in~\cite{Popov:1999} where Popov reinterpreted the results of the first solar axion helioscope search~\cite{Lazarus:1992ry} and announced a search at RICH~\cite{Baillon}, whose results were apparently never published. 
He also computed the solar HP flux from the eV to the keV region to interpret the null results as constraints on the parameter space, but did not publish these results. 
The solar flux of HPs in the keV range was presented in~\cite{Redondo:2008aa} where the low mass HP flux was found to be suppressed with respect to earlier calculations and the results of the CERN Solar Axion Search (CAST)~\cite{Zioutas:2004hi,Andriamonje:2007ew} were analysed and used to set constraints. 
However, the flux of {\em longitudinally} polarised HPs (disregarded in the previous studies) was severely underestimated for low mass HPs, as recently noted in~\cite{An:2013yfc} and confirmed in~\cite{Redondo:2013lna}. 
This opened up the possibility of detecting low mass HPs with the ionisation events of dark matter detectors such as XENON10~\cite{An:2013yua} by using the technique proposed in~\cite{Horvat:2012yv} (which by that time was thought to be inefficient because of the underestimated flux). A recent global fit of solar precision data (helioseismology and neutrino fluxes) set the strongest constraint on the solar flux of L-HPs and promises even tighter constraints if the solar abundance problem is eventually mitigated~\cite{Vinyoles:2015aba}.

\begin{figure}[tbp]
\begin{center}
\includegraphics[width=7.2cm]{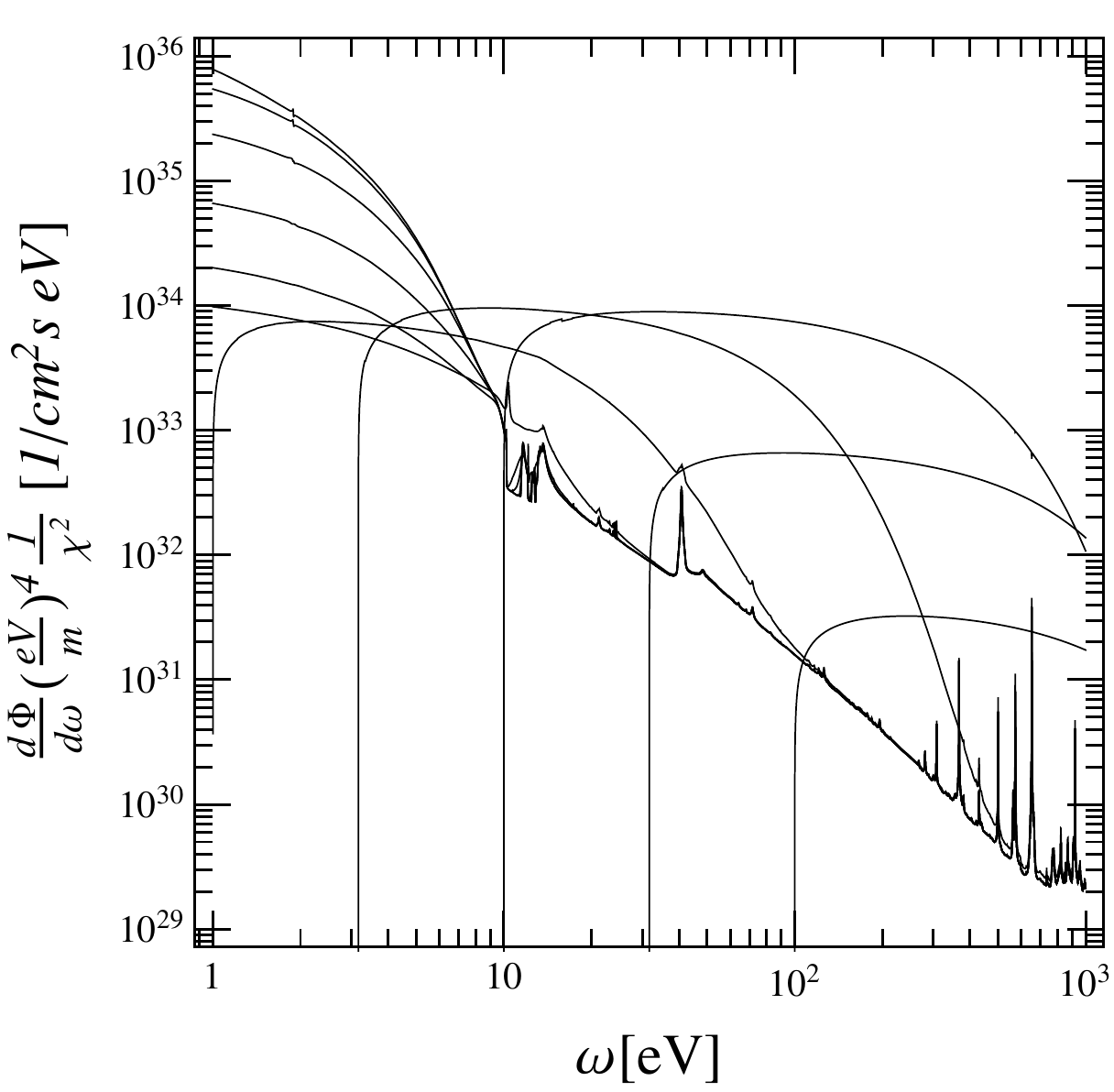}
\vspace{-.35cm}
\caption{\small Solar HP Flux as a function of frequency $\omega$ for a wide energy range for different HP masses from a 1-D solar model~\cite{Redondo:2014}. The highest six lines without threshold correspond to $m=10^{-3},3.16\times 10^{-3},10^{-2},3.16\times 10^{-2},10^{-1},0.316$ eV from top to bottom. The rest is recognisable through the threshold, $m=1,3.16,10,31.6,100,316$ eV. Note that the flux is normalised by a factor $\chi^2 m ^4$ in units of eV$^4$. We advance that in the frequency range $\omega\sim 3$ eV relevant for SHIPS and only for the lowest masses, this 1-D model flux is uncertain by O(1) factors, so we will use a 3-D model (Fig.~\ref{solarflux}) for the analysis of this paper.}
\label{solarflux1D}
\end{center}
\vspace{-.7cm}
\end{figure}

The most recent atlas of solar HP emission~\cite{Redondo:2014} provides for the first time an overall picture of the flux as a function of HP mass and frequency. The {\em transversely polarised} HP fluxes from a 1-D solar model for a wide frequency range are plotted in Fig.~\ref{solarflux1D}. For low mass HPs, the solar flux originates mostly from regions slightly inside the photosphere, and peaks in the visible and infrared. The high fluxes and the existence of suitable low background detectors make the visible range a very good option for a solar HP search. New experiments were recently proposed to search for solar HPs in the visible~\cite{Gninenko:2008pz}. Indeed, the CAST~\cite{oai:arXiv.org:0809.4581,Cantatore:2010zzb} and SUMICO~\cite{Mizumoto:2013jy,Inoue:2013jpa} collaborations already performed piggyback searches in the visible by adapting photomultipliers to their axion helioscopes.

In this paper we report on the results of a helioscope experiment to detect the solar flux of (transversely polarised) HPs in the visible energy range. Our instrument is based on the HP helioscope proposed in~\cite{Gninenko:2008pz}, which builds on the idea of Sikivie's axion helioscope~\cite{Sikivie:1983ip}. 
We describe our apparatus in section \ref{sec:exp} and our measurements in section \ref{sec:meas}. 
We analyse the results and implications in section \ref{sec:anal} and give our conclusions and outlook in section \ref{sec:conclu}.

%\newpage

\section{Experiment}\label{sec:exp}

\subsection{Concept}

\begin{figure}[t]
\centering
\includegraphics[width=155mm]{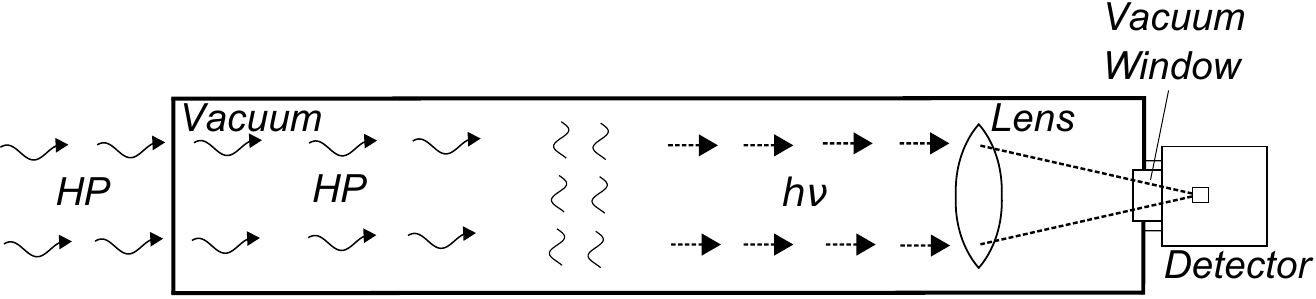}
\caption{Schematic overview of the oscillation process from HP to photon inside the SHIPS vacuum tube: The generated photons are collected by a Fresnel lens, focused onto a photo detector and counted.}
\label{scheme0}
\end{figure}

The Sun would be a copious source of HPs. If they have small sub-eV masses, the flux peaks in the infrared and below, with a sizeable amount still at visible energies~\cite{Redondo:2014}, see Fig.~\ref{solarflux1D}.  
A helioscope experiment consists of a light-tight chamber where HPs can enter and oscillate into photons that are to be detected. The light-tightness ensures that no other solar or ambient photons can enter the experiment and give false signals.  
When a relativistic HP passes trough the opaque shielding of the chamber, it has a probability to appear as a photon after a length $L$ behind the barrier given by\footnote{See~\cite{Okun:1982xi,Ahlers:2007rd} and references therein.} 
\be
\label{proba1}
P(\gamma\leftrightarrow {\rm HP})_{\rm vacuum}\simeq {4\chi^2} \sin^2 \left(   \frac{m^2} {4\omega} L \right),  
\ee
where $\omega$ is the photon/HP energy and $m$ the HP mass\footnote{We work in natural units, $\hbar=c=1$, where 1 meter = $5.07 \cdot 10^6 \ eV^{-1}$.}. This expression is valid for vacuum conditions (refraction index $n$=1). We discuss below the negative effect of residual gas in the interior zone of the chamber and give the maximal acceptable gas pressure. The photons have the same energy and propagation direction as the original HPs, so they indeed appear as coming from the exact place in the Sun where the HPs were produced\footnote{The atmosphere affects light propagation and to some extent it will also affect HPs. 
The HP index of refraction is $n_{\rm HP}\sim 1-m^2/(2\omega^2)+\chi^2(n-1)$, see~\cite{Jaeckel:2013eha} and assume that the HP mass term dominates in the atmosphere. 
Thus, atmospheric distortions are similar to the photonic ones (same sign) but a factor $\chi^2$ smaller, which makes them negligible.}.  
These photons can be focused with a lens onto a low noise photon detector. 
A schematic of a helioscope experiment, such as the one described in this paper, is depicted in Fig.~\ref{scheme0}. 

\begin{figure}[t]
\begin{center}
\includegraphics[width=8cm]{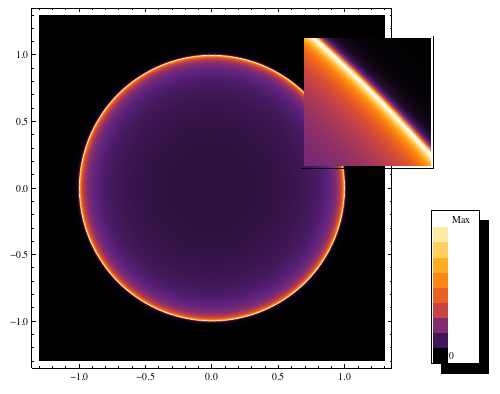}
\caption{Typical angular distribution of sub-eV mass HPs emitted from the Sun at visible energies $\omega\sim 3$ eV. The same distribution is expected for the photons arising from HP oscillations in an Helioscope chamber. Angles are normalised to the solar angular radius.}
\label{rim}
\end{center}
\end{figure}

Low mass HPs are mostly produced in the outer layers of the Sun - immediately inside the photosphere - so they fill the whole Sun's image, presenting a maximum of luminosity at a ring immediately below the solar surface~\cite{Redondo:2014}. The situation is very different to that of solar axions, which are mostly emitted with X-ray energies from the solar core~\cite{Redondo:2013wwa}. A typical HP angular distribution is depicted in Fig.~\ref{rim}. This mimics the photon distribution on the focal plane detector. 

The expected photon detection rate is tiny, as can be seen from the already existing bounds on kinetic mixing shown in Fig.~\ref{bounds}.
As an example, consider $\chi= 3\times 10^{-7}$ and $m=0.6$ meV, close to the exclusion limit of the ALPS experiment~\cite{Ehret:2010mh}. Then, the differential HP flux in the visible is large  
 
$\sim 2.3\times 10^{9}/{\rm cm^{2}s\, eV}$ (from Fig.~\ref{solarflux1D}), but the oscillation probability is tiny,  
$P\sim 3.6 \times 10^{-13}\sin^2(...)$, so that the differential photon flux is, at most $\sim {\cal O}$(mHz/cm$^2$eV), if the $\sin^2$ is of order 1. 
 
The only experimental parameter that we can use to maximise the $\sin^2$ factor, i.e. the oscillation probability, is the chamber length $L$, i.e. the distance between the chamber walls and the position of the collecting lens. Ideally, we want to make our experiment long enough to host at least one oscillation for the HP masses of interest, because for $L\ll 2\pi\omega/m^2\equiv L_{\rm osc}$ the probability is suppressed with respect to the maximum value ($4\chi^2$) by a factor $\propto (L/L_{\rm osc})^2$. $L_{\rm osc}$ depends on the a priori unknown HP mass. It is hard to beat the sensitivity of experiments searching for deviations of Coulomb's law and distortions of the CMB for masses below the benchmark $m_*=2\times 10^{-4}$ eV, and thus it is reasonable to target parameter space above it, i.e. $m>m_*$. 
Using a typical visible energy $\omega\sim 3$ eV we thus find that for $m>m_*$, $L_{\rm osc}< 0.1$ km and thus we would require $L>0.1$ km.  
This is quite a challenge indeed, but the closer one gets to this figure, the smaller will be the region of masses above $m_*$ for which the sensitivity is suppressed. 

Since our goal is to explore as much ($\chi,m$) parameter space as possible, the helioscope chamber length and collection area shall be as large as possible and the signal should be recorded for a long time to compensate the small flux.
The need of a large integration time suggests to track the Sun with the chamber and the optical system. 
A HP helioscope looks then very much like an ordinary telescope (with the lid on).    

With such small signals, the measurements are dominated by detector noise and parasitic backgrounds (radioactivity, cosmic rays, ...). The complementary measurements with the Sun far out of sight have to be undertaken to estimate the background level. Subtracting this `only background' from the `Sun' data (HP signal + background) will reveal the signal from solar HPs.

Note that the solar HP flux is maximal at the lowest energies but photo-detectors are also less sensitive at low energies. Working at visible energies $\omega\sim 3$eV is a good compromise between a sizeable flux, the availability of low dark-current detectors, simplicity and reliability of the set-up. 

\begin{figure}[bt]
\centering
\includegraphics[width=90mm]{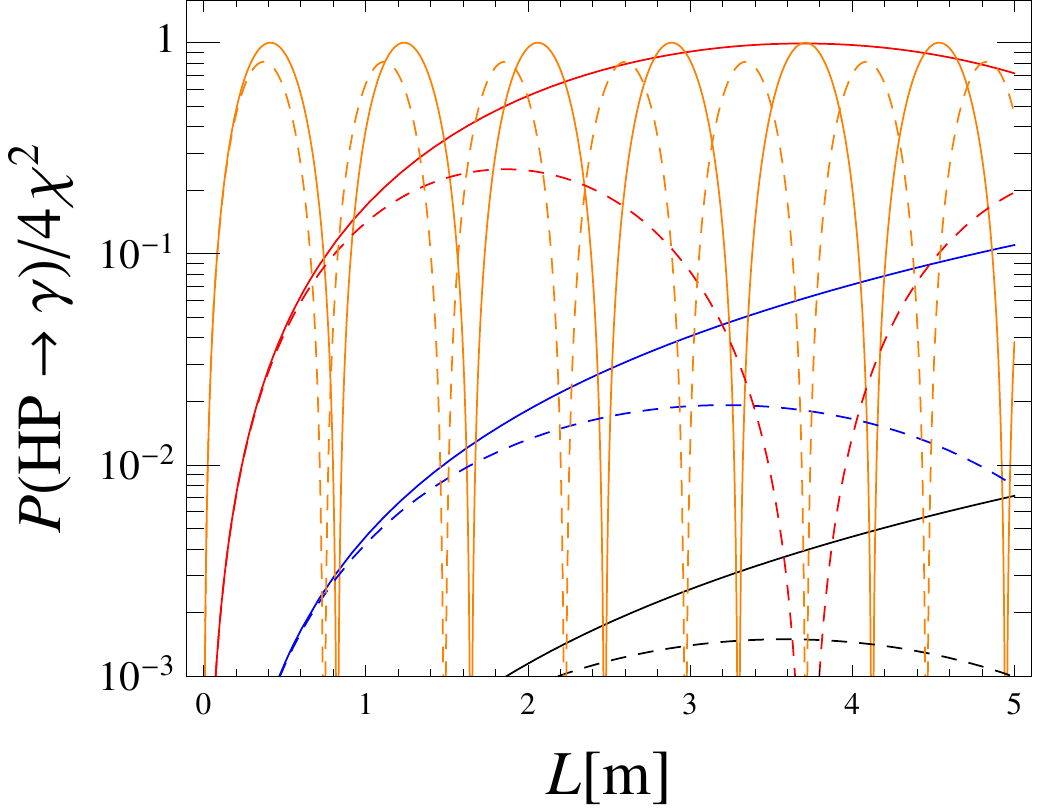}
\caption{HP $\to$ photon oscillation probability as a function of the length from an opaque wall. 
The HP/photon energy is $\omega=3$ eV. Different colours correspond to different HP masses: black,blue,red,orange for $0.2,0.4,1,3$ meV. 
Solid lines correspond to vacuum oscillations (pressure smaller than 0.001 mbar in this case) and dashed lines are for air at a pressure of 0.2 mbar and standard temperature conditions.  
 }
\label{oscila}
\end{figure}
 
Let us turn into the vacuum issue. 
The optimum estimates above correspond to HP$\leftrightarrow$photon oscillations in vacuum but, in realistic lab conditions, there will be always a bit of air inside the chamber. The presence of a medium modifies the photon propagation and this affects the photon$\leftrightarrow$HP oscillations, similar to neutrino oscillations.  
In a medium with an index of refraction $n=n(\omega)$ (assumed transparent) the oscillation probability modifies to 
\be
\label{proba}
P({\rm HP}\to \gamma)\simeq \frac {4\chi^2 m^4} {(m^2 - m^2_{\gamma})^2 + 4 \chi^2 m^4} \cdot \sin^2 \left(   \frac {\sqrt {(m^2 - m^2_{\gamma})^2 + 4 \chi^2 m^4}} {4\omega} L \right), 
\ee
where we have defined an {\rm effective photon mass}
$m_\gamma^2 = \omega^2(1-n^2)$. 

Air at visible energies (like all neutral gases) has $n(\omega)>1$, which makes $m_\gamma^2$ negative. 
The amplitude of the oscillations is now smaller than $4\chi^2$ for any finite value of $n>1$, suppressed by a factor
\be
\frac{ m^4}{(m^2+\omega^2(n^2(\omega)-1))^2+4\chi^2m^4}, 
\ee 
with respect to the vacuum case\footnote{Except at very small $L\ll L_{\rm osc}$ because the oscillation length $L_{\rm osc}$ decreases by the square root of this factor and thus compensates.}. The situation is exemplified in Fig.~\ref{oscila} where we plot the probability as a function of distance for different HP masses (diff. colours) in vacuum (solid lines) and with a significant gas pressure (dashed lines). 
For the lowest HP masses in vacuum (solid black, blue for $m=0.2,0.4$ meV) we see the suppression of the amplitude due to the fact that $L<L_{\rm osc}$: the probabilities do not reach $4\chi^2$ within 5 m. The dashed lines correspond to a very low air pressure of $0.2$ mbar, where we see a very significant suppression of $O(10^{-3},10^{-2})$ of the maximum, which now comes inside the 5 m available length. The amplitude of the oscillations is always equal or smaller than the vacuum case. 
Note also that for higher masses (red and orange for 1 and 3 meV, respectively) the suppression of the amplitude becomes more moderate.  

If in our experiment, we want to keep this medium suppression smaller than, for instance, a 10\% correction for all HP masses above $m_*$, we have to ensure that the index of refraction satisfies the following criterium,  
\be
n(\omega)-1 < 0.027 \frac{m_*^2}{\omega^2}. 
\ee
Using $n_{\rm air}-1\simeq 2.8\times 10^{-4}$, standard temperature and pressure conditions with ($p_{stp}=1$ atm, $T_{stp}=293.15$K) for\footnote{The dependence of $n$ in the visible-IR range is relatively small, so we omit it.} $\omega=1.2$ eV and the ideal gas law, the above condition translates onto an upper limit of the pressure in the oscillation region 
\be
p < 10^{-3} {\rm mbar} \(\frac{m_*}{10^{-4} \rm eV}\)^2\(\frac{1\, \rm eV}{\omega}\)^2 \frac{T_{stp}}{T}. 
\ee
If this inequality is respected, the probability will be given by the vacuum formula \eqref{proba1}. 

\subsection{Technical setup}

\begin{figure}[ht!]
\begin{center}
\subfigure[TSHIPS mounted piggyback on the OLT]{\includegraphics[width=70mm]{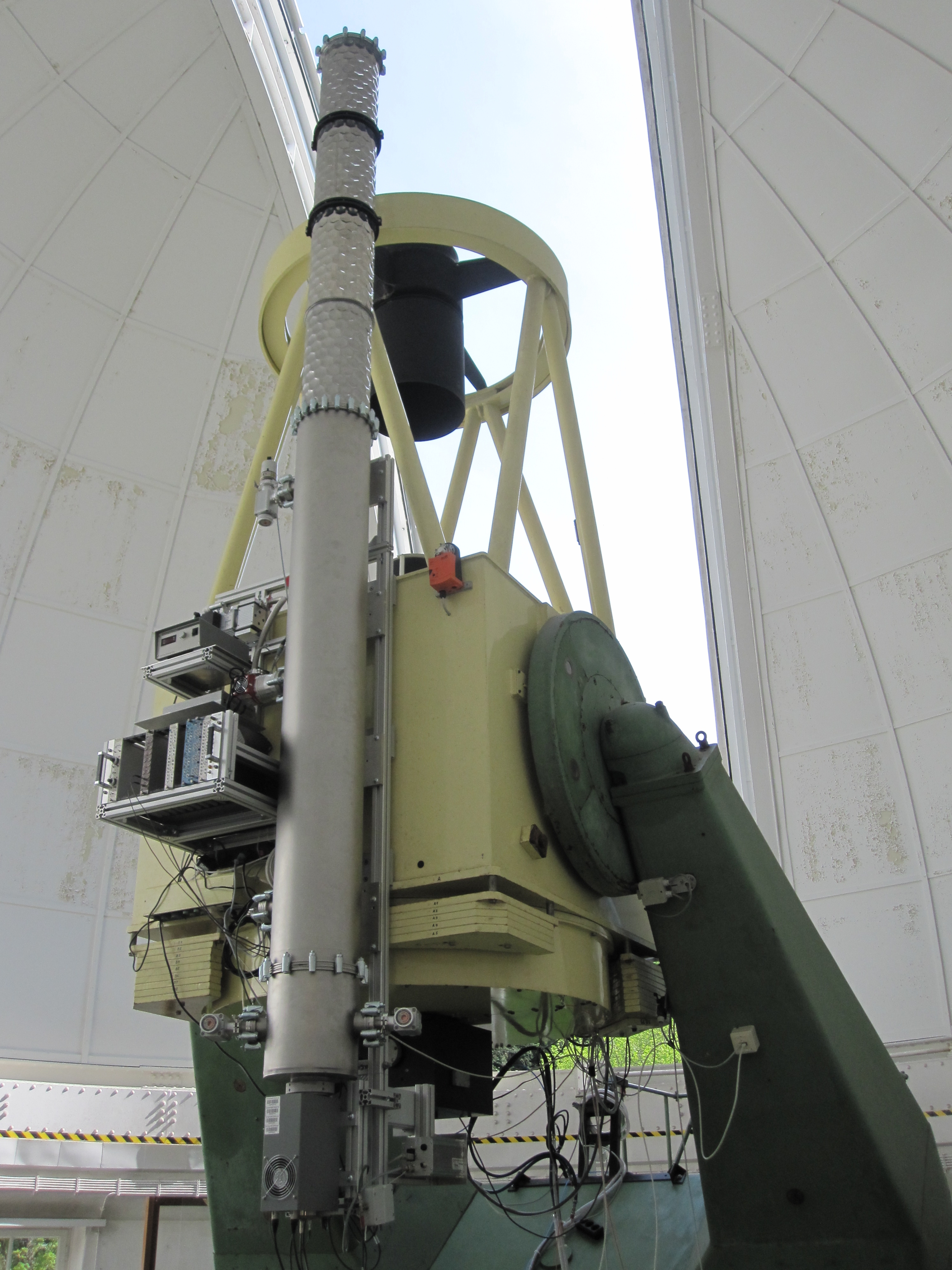}}
\subfigure[Schematics of TSHIPS]{\includegraphics[width=40mm]{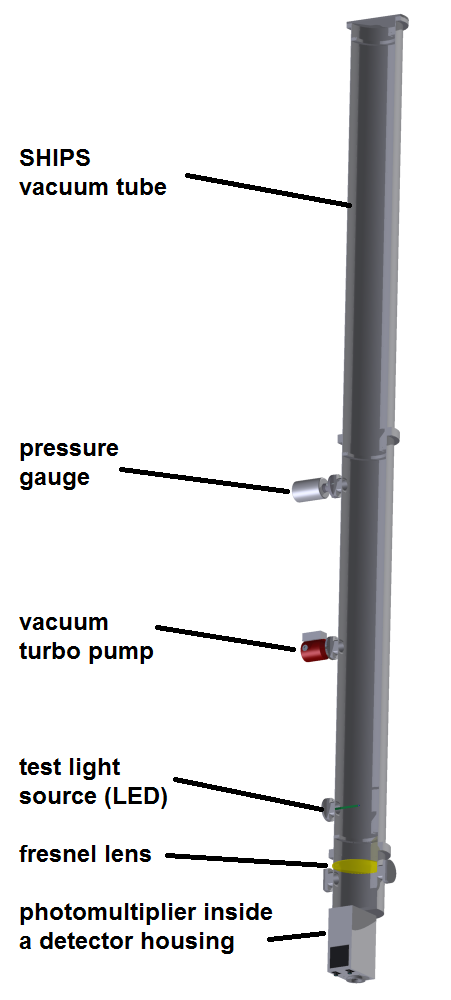}}
\caption{The Telescope for Solar Hidden Photon Search (TSHIPS) attached to the Oskar L\"uhning Telescope (OLT) at Hamburg-Bergedorf}
\label{TSHIPS}
\end{center}
\end{figure}

The Telescope for Solar Hidden Photon Search (TSHIPS) is located on the premises of the observatory in Hamburg-Bergedorf. It is comprised by the combination of two long tubes and an additional smaller prolongation in the bottom part, see Fig.~6 (a). The total length is 430 cm, the inner diameter $25.9\pm 0.2$ cm (depending on the position along the tube). The middle tube is a 200 cm long stainless steel tube to which the vacuum pumps and pressure gauge are attached.  
The upper tube is also 200 cm long and was developed for this project. It is a prototype lightweight vault structure, manufactured by the Dr. Mirtsch W\"olbstrukturierung GmbH in Berlin, with a wall thickness of just 0.8 mm and a weight of only 14.5 kg.  Nevertheless it provides the same stability, stiffness and vacuum properties as the middle tube with its 75 kg and 3 mm thickness. 
The small prolongation is used to house the optics and prepared for easily implementing further detectors and concepts. 
The detector used in the physics runs discussed in this paper is mounted on an independent housing below the small compartment. 
TSHIPS is mounted piggyback on a major telescope of the observatory, the so called Oskar L\"uhning Telescope (OLT), cf. Fig.~\ref{TSHIPS} a. 
The helioscope itself and the OLT - utilised as the mount for TSHIPS - are fully remote-controlled. The equatorial OLT mount has an operation range in azimuth of 360\deg and a 
range from 10\deg to 90\deg in altitude.

\begin{figure}[ht!]
\begin{center}
\includegraphics[width=6cm]{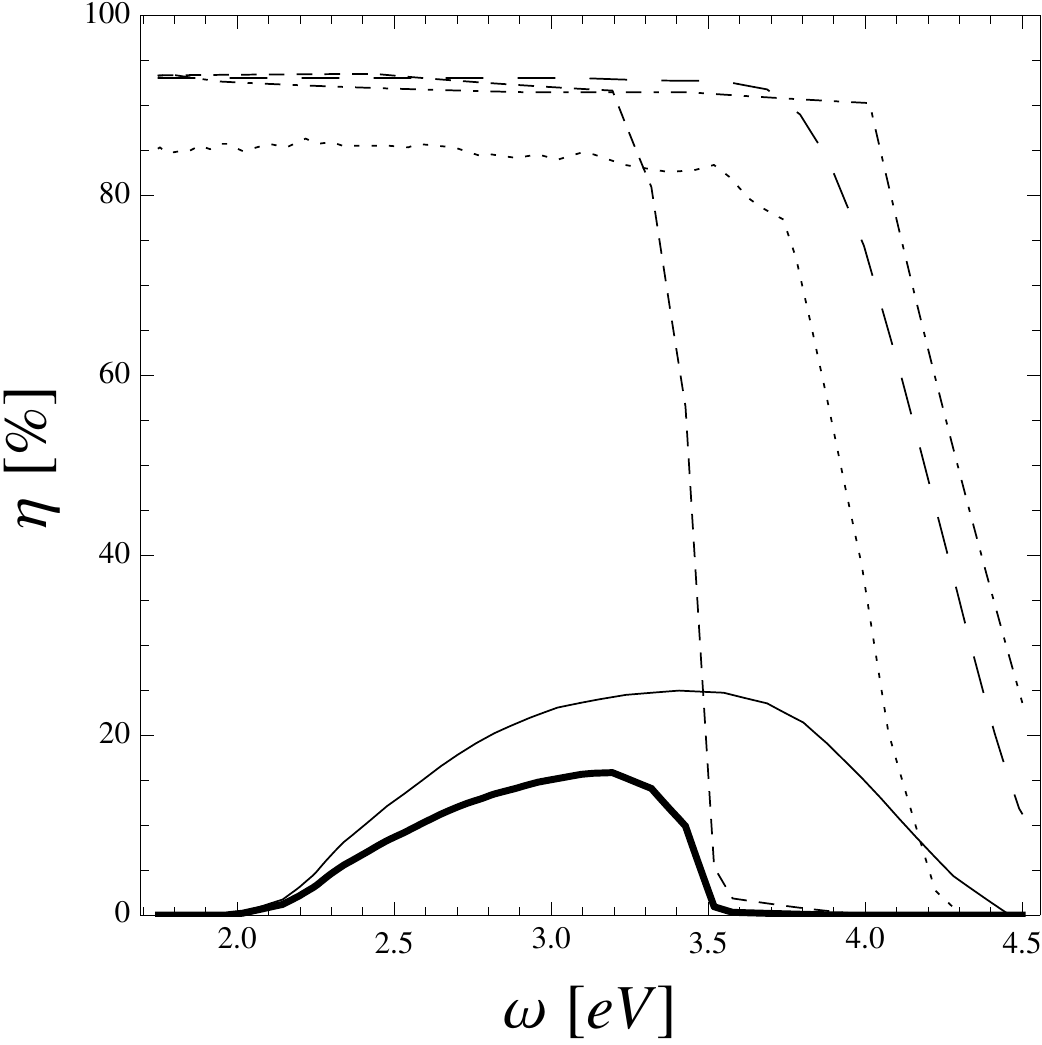}
\caption{Efficiency factors: quantum efficiency of the PMT ($\eta_{\rm PMT}$, thin solid line) and transmissivities of the PMT's front window ($\eta_{\rm 9893/350B}$, long dash), of the cooler housing window ($\eta_{\rm Fact50}$, dotted), of the vacuum window ($\eta_{\rm vw}$, dot-dashed), 
of the Fresnel lens ($\eta_{\rm Fres}$, short dash), and the total detection efficiency ($\eta$, thick solid)}
\label{efficiencies}
\end{center}
\end{figure}

Photons from HP oscillations inside TSHIPS are focused onto the detector by means of a acrylic Fresnel lens, qualified for observations in the optical and near infrared spectral range. The Fresnel optics fulfils all the requirements on the basis of transmission, $> 90\%$ (see $\eta_{\rm Fres}$ in Fig.~\ref{efficiencies}), light concentration and aberration for our non-imaging solar PMT observations. The lens' optical active area has a diameter of 25.4 cm and the focal length is 20.32 cm. 
The Sun's image (angular size of 0.53\deg) in the focal plane has a diameter of $1.9$ mm.  
The lens is held in place inside the small compartment by a bipartite lens holder made of aluminium rings, see sketch in Fig.~\ref{TSHIPS} (b).  
The first part defines the principal site of the holder and gets attached directly to the inner wall of TSHIPS\footnote{The position of the lens inside the compartment can be adjusted over a wide range in order to accommodate for different detectors and their varying active area positions. }. 
The second and slightly smaller one carries the lens and is connected to the first by means of three adjustable screws with cup springs in equidistant distributions, which allows us to align the lens to the optical axis and the detector. 
 
The focal point of the lens is adjusted to be right outside the vacuum vessel, coinciding with the photomultipliers's photocathode. A transition window conducts the focused photons through the vacuum seal right onto the detector. The transmissivity of the vacuum window, $\eta_{\rm vw}$ is shown in Fig.~\ref{efficiencies}.

Proper vacuum conditions are indispensable for successful data-takings with TSHIPS to ensure unsuppressed HP generation rates.
Permanent pre-vacuum of 10$^{-2}$ mbar in normal operation mode is created and maintained by a membrane pump. 
During data taking, a turbopump directly attached to TSHIPS - see Fig.~\ref{TSHIPS} (b) - establishes a pressure smaller than 10$^{-4}$ mbar in the helioscope volume of more than 260 litres within minutes. 
The pressure inside the tube can be measured by a (Pfeiffer Vacuum PKR 251) pressure gauge 
attached directly to the conversion region of the helioscope in the middle of the tube.

As photon detector, we employed an ET Enterprises 9893/350B photomultiplier (PMT) tube for all the solar HP observations. 
The PMT features a 9-mm diameter photocathode with good quantum efficiency in the violet-blue-green region (peak quantum efficiency of 25\% at $\omega \sim 3.5$ eV), see Fig.~\ref{efficiencies}. It covers generously the 1.9 mm size of the Sun's image. The photocathode is in a vacuum tube protected by a borosilicate window with good transmissivity at the frequencies of interest, see $\eta_{9893/350B}$ in Fig.~\ref{efficiencies}. 
 
Photon events in the PMT produce characteristic output current pulses which are amplified and then recorded by a DRS4 Evaluation Board~\cite{DRS}. This board is equivalent to a digital oscilloscope and recognises, records and stores every single event (current pulse). The DRS4 was operated with 5 giga-samples per second and performed very stably and accurately. 
We keep the operating voltage of the PMT fixed at 2.2 kV during the observation phase to avoid errors in the justification of the voltage. In this regard, exclusively the errors in the voltage output of the power supply itself occur as variations in the measured fluxes. 
With the minimisation of drift effects and other variations in the dark count rate of the PMT one prevents a major systematic error at the same time.
A major source of dark count rate fluctuations originates from temperature variations \cite{Hamamatsu:2007}. To reduce them to a minimum, the PMT is placed in a ET FACT50 cooler housing which keeps the temperature at -21\deg C by means of self-regulated cooling power. Photons enter the cooler housing through another window, with good transmissivity in the visible and infrared, see $\eta_{\rm Fact50}$ in Fig.~\ref{efficiencies}. 
We recorded the ambient temperature and humidity in the helioscope dome during the whole data taking period for controlling purposes and found that they are - as expected due to the self-regulated cooling - fully uncorrelated with the count rates.

An all-round detector interface was designed for TSHIPS to be able to utilise different types and models of detectors beside PMTs like charged coupled devices (CCDs) with modifications as easily, trouble-free and safe as possible. TSHIPS is equipped with several valves for any potential further devices.

Inside the tube, a device with a blue and red LED adjustable in brightness and flashing frequency serves as a test source for detector gauging purposes. 
Among other measures the LEDs were dimmed to very low fluxes to verify and guarantee the sensitivity of the PMT in the photon counting mode.

The light tightness of the whole instrument was secured by longtime tests at different times, inner pressures and temperatures to cover different conditions of the surrounding light irradiation fields, deformations and degrees of thermal expansion of the parts of the tube. The PMT mounted on the instrument finally showed general constant dark count rates $\sim$ 0.46 Hz agreeing with the stated value of the manufacturer and scientific analyses like \cite{Lozza}. 

In order to test the precision and long-term stability of the Sun tracking with TSHIPS the set-up was converted to an ordinary telescope by replacing the upper steel cover of the upper tube with a transparent flange carrying a solar filter and mounting an Apogee U4000 CCD camera at the detector interface. The pointing and tracking precision and stability was accurately determined by monitoring the solar disc on the Apogee's chip. Its shift is directly proportional to the tracking errors of the helioscope. The Sun's position drifted in a whole day by $7.2'$, much smaller than the 2.5\deg angular acceptance of our 9mm diameter PMT. Hence pointing and longterm tracking of TSHIPS were verified as adequate for unproblematic measurements. Beside this, the marginal motion of the position of the Sun spot on the photocathode of the PMT does not influence in terms of sensitivity.

\section{Measurements}\label{sec:meas}

The data for this publication was obtained in photon counting mode of our 9893/350B photomultiplier tube, whose pulses were registered and stored by the DRS4 Evaluation Board. The final data consists of 660 net hours of measurements taken with TSHIPS between the 18th of March and the 7th of May of 2013. 

\begin{figure}[ht!]
\begin{centering}
\makebox[\textwidth][c]{\includegraphics[width=155mm]{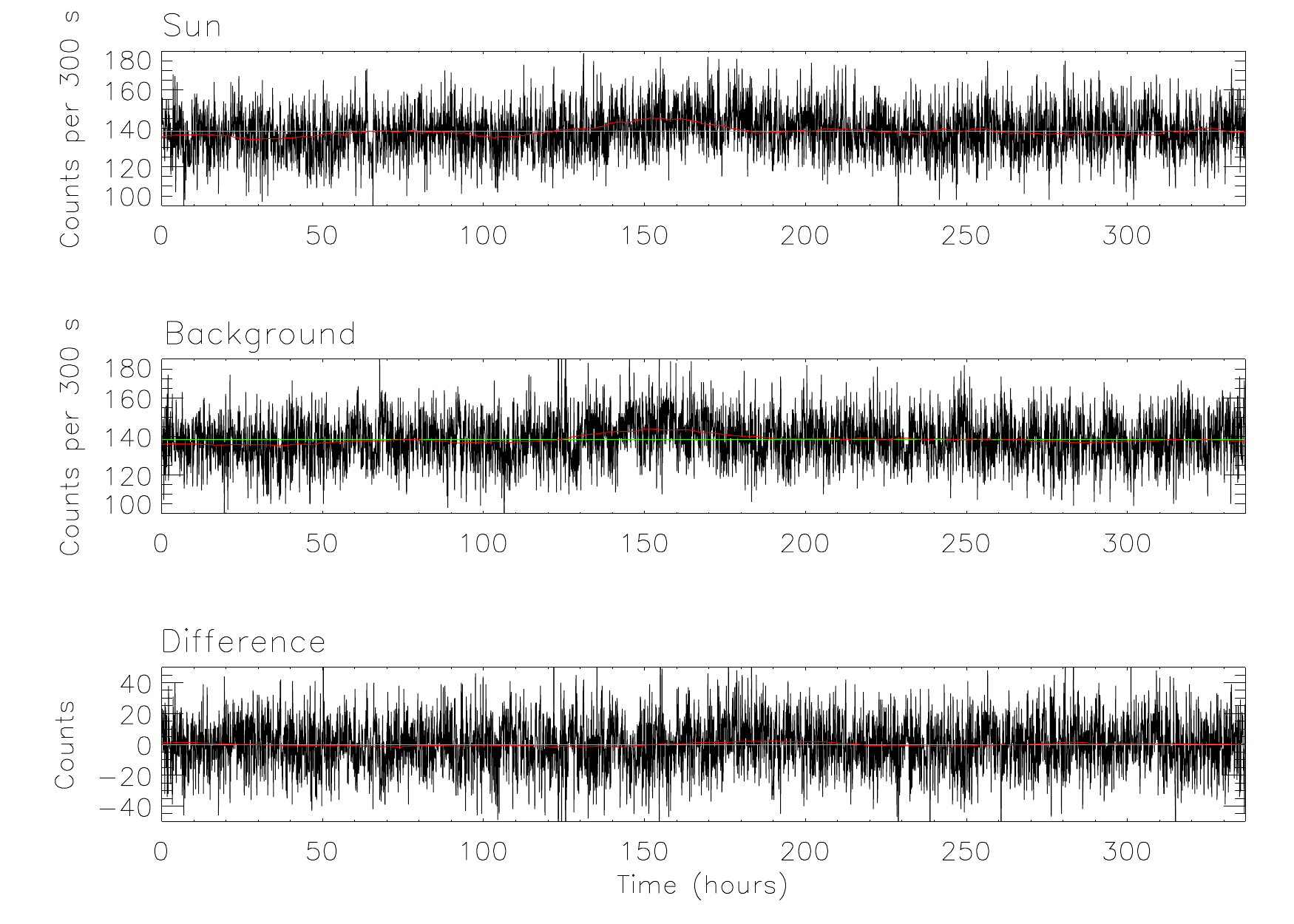}}
\caption{Comparison of the Sun (top) and background (centre) measurements as well as the differences between them (bottom). All plots show the corrected counts of events of subsequent 5 minute intervals as function of time. Subsequent measurements have been merged to produce a continuous sequence. The black lines represent the data, the green ones show the mean values and the red curves are smoothed fits to the data.}
\label{plot:multiplot}
\end{centering}
\end{figure}

We recorded the PMT photon events while tracking the Sun with TSHIPS during 330 hours. 
Each day, the Sun was tracked above the horizon for a period of time symmetric around midday. 
This data was divided in 5 minute time-frames (a total of 4041). The count rate of this `Sun' data is shown in Fig.~\ref{plot:multiplot} (top) as a function of the measurement time. In this `Sun' data set we expect photons from solar HP conversion plus general background. 

\begin{figure}[ht!]
\begin{centering}
\makebox[\textwidth][c]{\includegraphics[width=125mm]{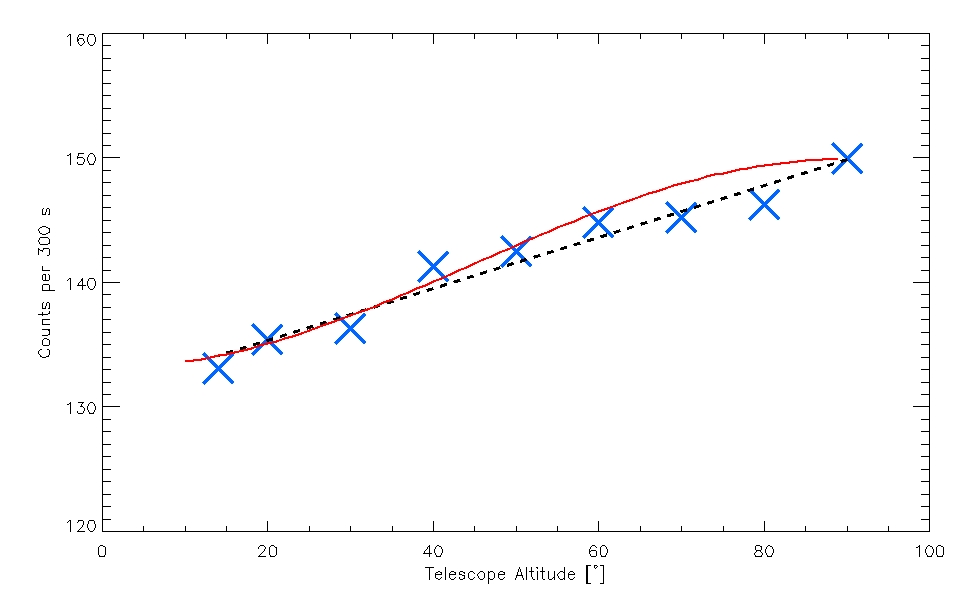}}
\caption{SHIPS PMT count rate as a function of the zenithal angle showing a dependence with the cosmic muon flux (blue crosses). Shown also is a linear fit (black dotted line) and another fit (red line) to a $\cos^2\theta$ dependence to the altitude-dependent part.}
\label{alt}
\end{centering}
\end{figure}

In order to reveal the events due to solar HPs we needed to measure a background set independently that reproduced as accurately as possible the background events we record while tracking the Sun. 
Typical noise contributions to PMTs' counting rates beside the emission of thermal electrons are radioactivity and cosmic rays~\cite{Hamamatsu:1998}. Both might depend on the position and azimuth of TSHIPS.   
In particular, cosmic muons\footnote{At sea level 98 $\%$ of cosmic rays are muons~\cite{cosmic rays}} induce Cherenkov photons in the Fresnel lens and the various windows, and their flux is $\propto \cos^2\theta$, where $\theta$ is the zenithal angle. 
Thus, the muon-induced background events depend on the solar track of one particular day, and a sort of azimuth average. In Fig.~\ref{alt} we show the PMT count rate as a function of the altitude of the TSHIPS pointing angle for a small data set. The data (blue crosses) fit well to a background base level of 135 counts/300 s (0.45 Hz) plus a $\cos^2\theta$-dependent contribution of 15 counts/300 s (0.05 Hz) at zenith. This seems to agree well with rough estimates of Cherenkov flux from the windows\footnote{The angles-integrated flux of cosmic muons above 1 GeV energies is $\sim 1/($cm$^2$ min) ~\cite{cosmic rays2} and produces around 20 visible Cherenkov photons/muon in a mm-thick Silicon window. Assuming $\sim O(10\%)$ detection efficiencies leads to the observed magnitude.   
}.

With this in mind, we recorded PMT events during night while performing {\em the same solar tracking we did during the day}\footnote{The alternatively considered method - to take Sun and Background measurements alternatively in short time-steps by pointing the helioscope towards and away from the Sun during the day tracking - would have been disadvantageous due to technical and other reasons.}.  
Thus, `Sun' and `Background' datasets were recorded as equivalent pairs of the exact same combinations of altitude, horizontal orientation respect to the lab, and further spatial ambient conditions. Note that this also minimises the potential influences on the data of variations in surrounding electromagnetic fields, geomagnetic fields or any other directional interferences at varying orientations of TSHIPS on the PMT signal. 
If it was not possible to complete this approach for both runs, maybe due to the OLT being used for other astronomical observations at good nightly weather conditions, the data previously taken from the Sun measurement was eliminated from the database and the recording of a new pair was started. 
The 330 hours of the `Background' are also divided in 5 minute frames and presented in 
Fig.~\ref{plot:multiplot} (centre). 

The difference between the count rates is proportional to the HP signal and is presented in Fig.~\ref{plot:multiplot} (bottom). The red line shows a smoothed count rate over a few hours for the two data sets and the difference and the green lines show the mean values during the whole data taking period. 

As we will see in the analysis chapter, the count rate for both data sets is very well described by Gaussian distributions with quite constant averages of $\sim 140$ counts/5 min $\sim0.46$ Hz with a standard deviation of just 0.039 Hz.
There are no signs from influences like day-night cycle, temperature, random fluctuations of operating voltage which would lead to anomalies in the Gaussian statistics.  

Despite being able to reduce general and directional background interferences and normally stable background rates even on scales of days and longer, we can identify a systematic upwards fluctuation {\em parallel in both datasets} for about 3 days. In Fig.~8 (top and middle) we can see this around hour 160 by a bulge in the red line that indicates the smoothed mean count rates. The origin of this temporary increase of the background is unknown. These fluctuations were not correlated to any of the mentioned monitored variables. Further on we found that this residual does not influence the difference between both data sets. This suggests that our approach of data taking strongly helps to prevent influences also from such possible systematics.   

At every data taking run, the required pressure conditions had to be established to avoid suppressing the HP oscillation probability.
It was not possible to monitor the pressure of the vessel during data taking since the pressure gauge operates using residual gas ionisation, which has an impact on the count rate of the PMT. Hence the pressure was checked to be below the required value of 10$^{-4}$ mbar regularly in between data taking lapses. 
Actually, we found every single pressure in the order of 10$^{-6}$ mbar in all our checks. Since the pressure was at any time two orders below the desired value it is appropriate to exclude an impact on the oscillation rate as a systematic error here.  

\section{Analysis and Results}\label{sec:anal}
\label{analyses}

\subsection*{Data reduction}

\begin{figure}[ht!]
\begin{center}
\subfigure[Voltage curve of the peak pulse of a typical PMT event]{\includegraphics[width=70mm]{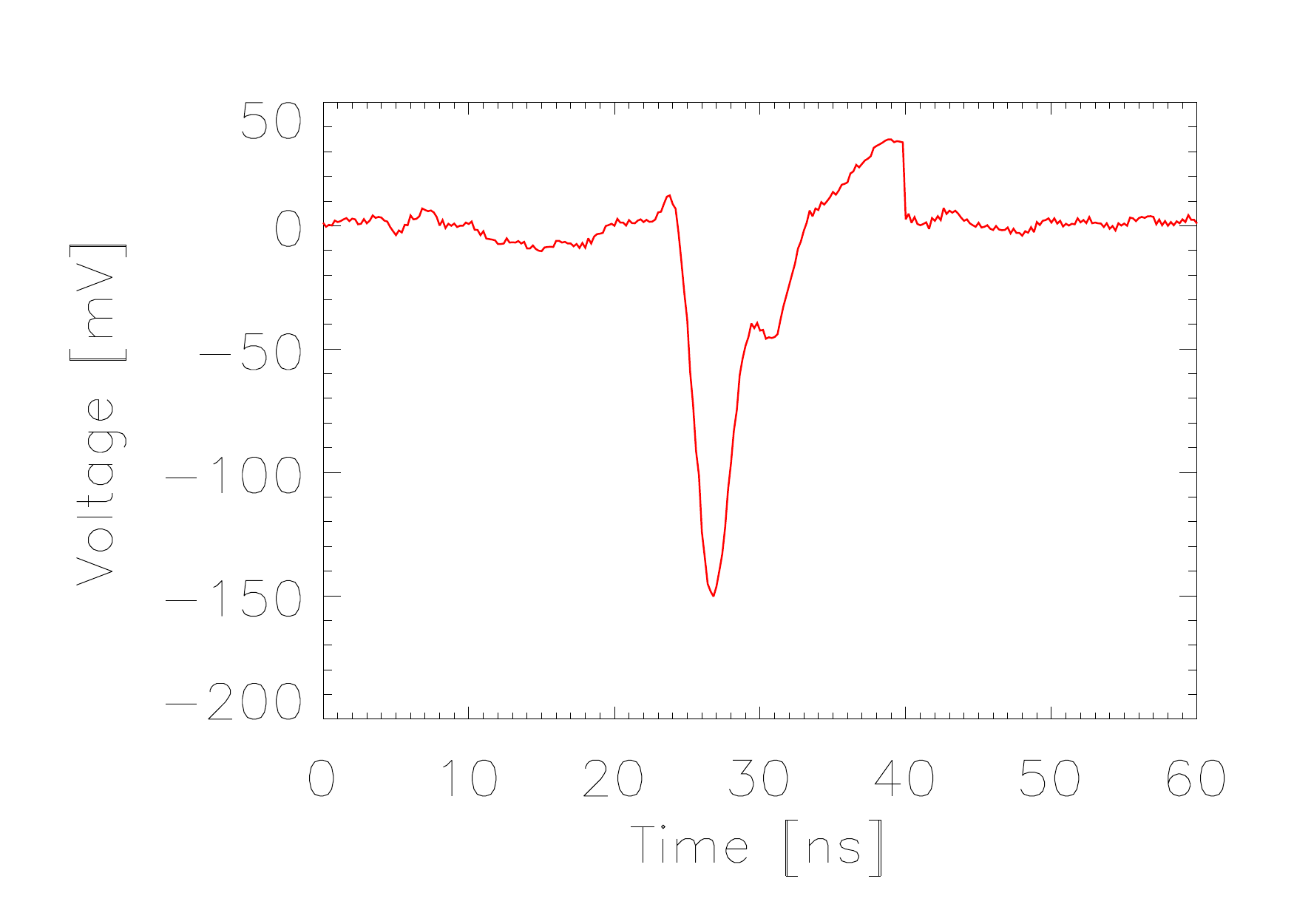}}
\subfigure[Stringing together of several signal pulse amplitudes each with their surrounding 1000 data bins]{\includegraphics[width=70mm]{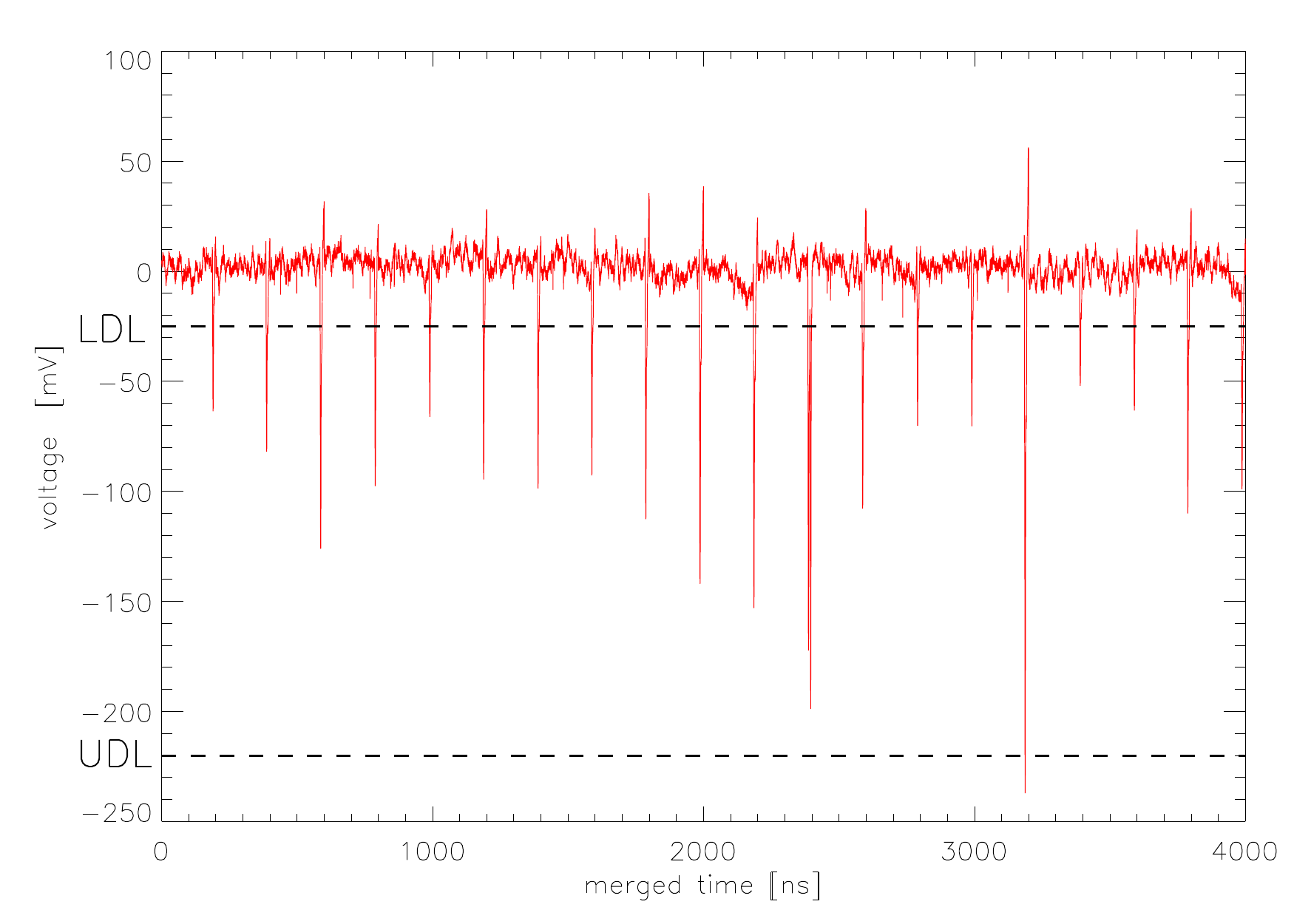}}
\caption{Illustration of the recorded voltage over the data bins according to triggered signals}
\label{plot:scheme}
\end{center}
\end{figure}
 
The recorded PMT current peaks show all the same characteristic shape but different amplitudes. A typical voltage peak from our signal and the according baseline is shown in Fig.~\ref{plot:scheme} (a).

Pulses with very low amplitudes can be interpreted as electronic noise, while large amplitudes are most likely induced by highly energetic particles induced by radioactivity or cosmic rays~\cite{Hamamatsu:1998}. With a lower and an upper discrimination level we can dismiss these events in our measurements as noise as it is the general approach in the single photon counting mode.
Following its standard analysis routines \cite{Hamamatsu:1998} we can define a window from about 25 to 220 mV where pulses are mainly due to photons. We thus define candidate photon events as pulses in this range and dismiss all events below and above these discrimination levels.  
A larger time window showing more pulses is displayed in Fig.~\ref{plot:scheme}  (b) to illustrate the varying pulse amplitudes of the recorded events. Shown is a sequence of 21 pulses with their surrounding data bins and the set discrimination levels. 

\subsection*{Data analysis}

\begin{figure}[ht!]
\centering
\includegraphics[width=150mm]{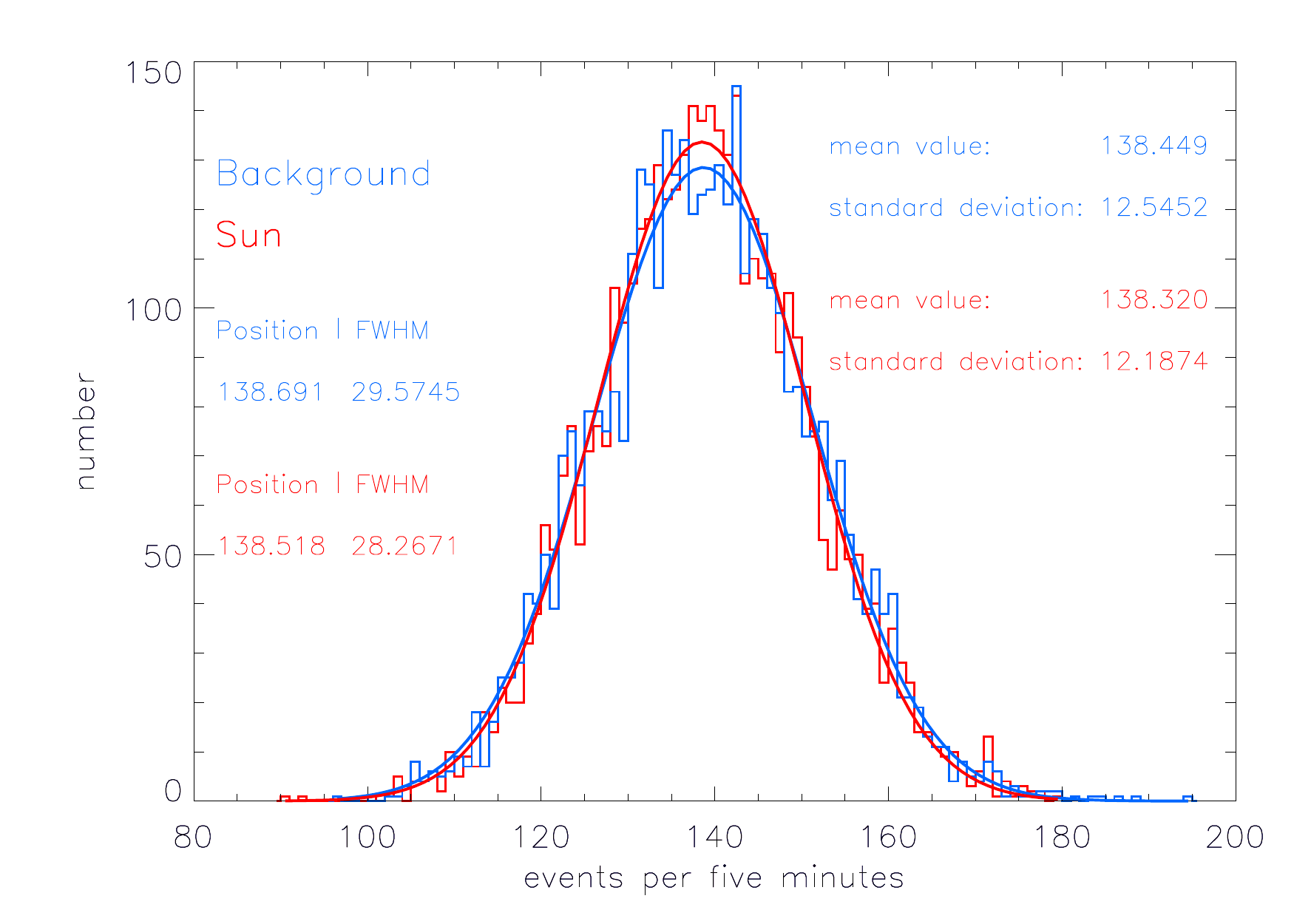}
\caption{Distribution of signal+background events per 5 minutes (Sun/red) and according Background distribution (blue). Both follow a Poisson distribution and are nicely fitted by very similar Gaussian distributions due to their high amount of events. The fitted mean values are ($138.52 \pm 0.20$) for the Sun and ($138.69 \pm 0.17$) for the background data sets.}
\label{overflow}
\end{figure}

The number of candidate event pulses was counted in each of the single 4041 5-minute data set pairs - separately for `Sun' and `Background' data sets. 
The resulting distributions of the time-binned events are shown in Fig.~\ref{overflow}. 
The `Sun' data set contains signal+background events and its distribution is shown in red colour while the only-background distribution is displayed in blue. 
A solar HP flux would reveal itself as a positive shift between the mean values of the `Sun' and `Background' distributions. 
However, the mean values of the `Sun' data set, $138.52 \pm 0.20\equiv (0.46173\pm0.00067)$Hz, and the only-background data set, $138.69 \pm 0.17\equiv (0.46230\pm 0.00057)$Hz, agree within their statistical errors. We do not find any evidence of a solar flux of transversely polarised HPs.

The errors are well understood by Poissonian statistical noise $\sigma/\sqrt{N}\sim 12.35/\sqrt{4041}\sim 0.20$ (see Fig.~\ref{overflow}) pointing towards a statistics-limited measurement. 

\subsection*{Results}

From the small difference of the means of the data sets (-0.173) compared to its standard deviation ($\sqrt{0.20^2+0.17^2}=0.26$) TSHIPS cannot provide evidence for the existence of hidden photon induced events but can constrain their appearance. 
Using the method of Feldman and Cousins we can build upper limits on the rate of detection of photons from solar HP oscillations,

\bea
\frac{dN}{dt} < 1.2\; \rm mHz, \quad at \quad 95\% C.L., \\
\frac{dN}{dt} < 1.8\; \rm mHz, \quad at \quad 99\% C.L., 
\eea 
and the photon fluxes inside TSHIPS, 
\bea
\frac{dN}{dt dA} < 25\; \frac{\textnormal{mHz}}{\textnormal{m}^2} \quad \text{at  \quad 95\%  C.L.},  \\
\frac{dN}{dt dA} < 36\; \frac{\textnormal{mHz}}{\textnormal{m}^2} \quad \text{at \quad 99\% C.L.,}
\eea 
respectively.

\begin{figure}[t]
\centering
\includegraphics[width=70mm]{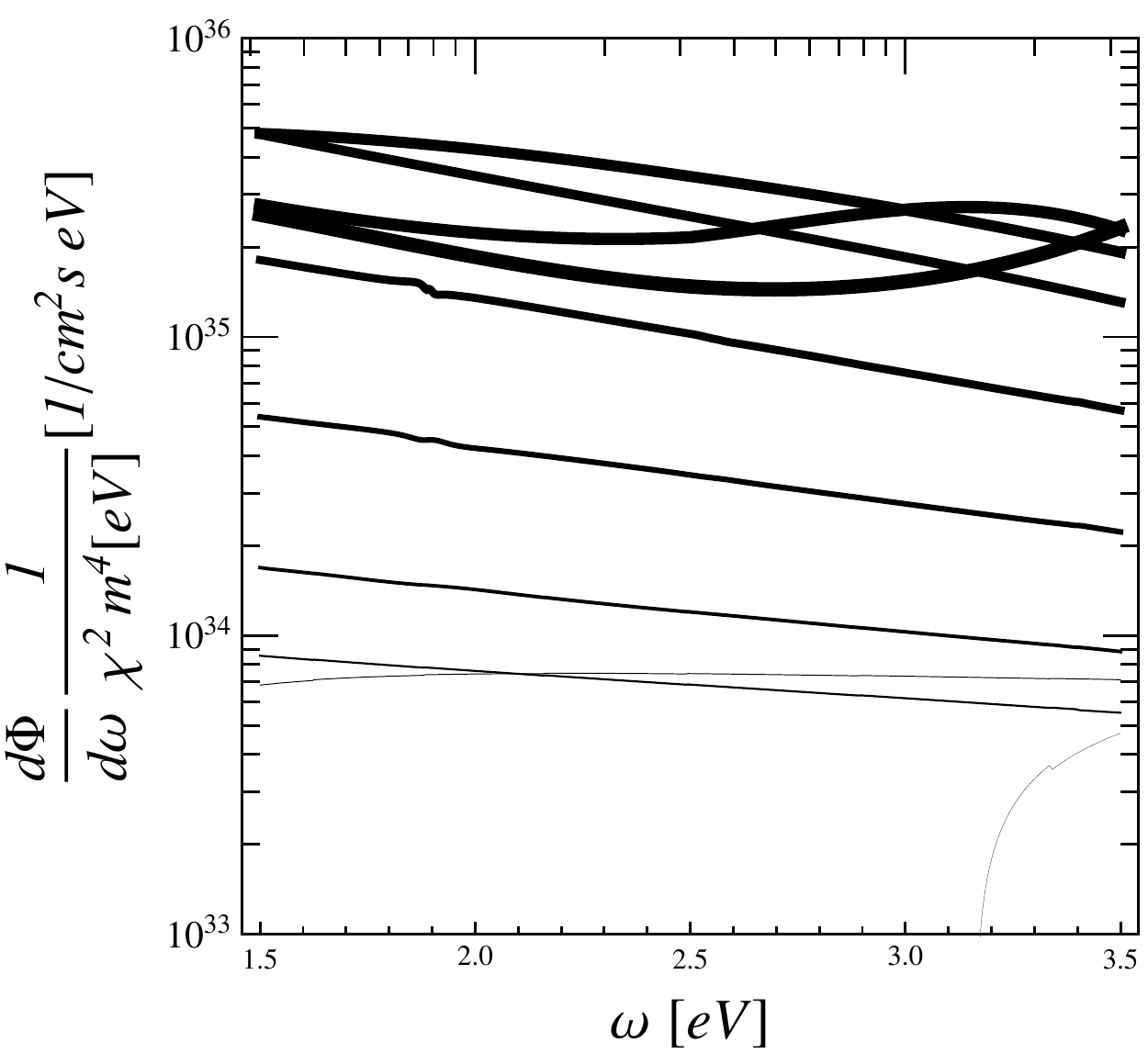}
\caption{Solar flux of hidden photons in HPs/(cm$ ^{2}\cdot$s$\cdot$eV) divided by the factor $\chi^2(m/{\rm eV})^4$ as a function of the HP energy in our region of interest. 
The curves correspond to masses $m=3.16,1,0.316,0.1,0.0316,0.01,0.005,0.003,0.002,0.001$ eV from the thinnest to the thickest. 
The data is taken from Ref.~\cite{Redondo:2014}.} 
\label{solarflux}
\end{figure}

We can translate our exclusion limit into a constraint on the HP parameters. The expected detection rate of photons from solar HP oscillations in TSHIPS is   
\be
\frac{dN}{dt} = A  \int \frac{d\Phi_{\rm HP}}{d\omega}
 P({{\rm HP} \rightarrow \gamma}) \eta(\omega) d\omega
\ee
where $A$ stands for the aperture area of TSHIPS, $A=499$ cm$^2$, $\frac{d\Phi_{\rm HP}}{d\omega}$ is the differential solar HP flux (function of $\chi$, $m$ and $\omega$) shown in Fig.~\ref{solarflux}, $P({{\rm HP} \rightarrow \gamma})$ is the oscillation probability given by \eqref{proba1} and
the overall detection efficiency $\eta(\omega)$ is given by
 
\be
\eta(\omega)=\eta_{\rm Fact50}\cdot \eta_{\rm 9893/350B}\cdot \eta_{\rm vw}\cdot \eta_{\rm Fres}\cdot Q_{\rm PMT}, 
\ee
where $ \eta_{\rm Fact50}$, $ \eta_{\rm 9893/350B}$, $ \eta_{\rm vw}$ and $ \eta_{\rm Fres}$ are the transmissivities of the 
  front window of the Fact50 cooler housing, 
  the  window of the 9893/350B, 
  the vacuum window 
  and the Fresnel lens, and 
$Q_{\rm PMT}$ the quantum efficiency of the PMT, see Fig.~\ref{efficiencies}. 

\begin{figure}[h]
\centering
\includegraphics[width=90mm]{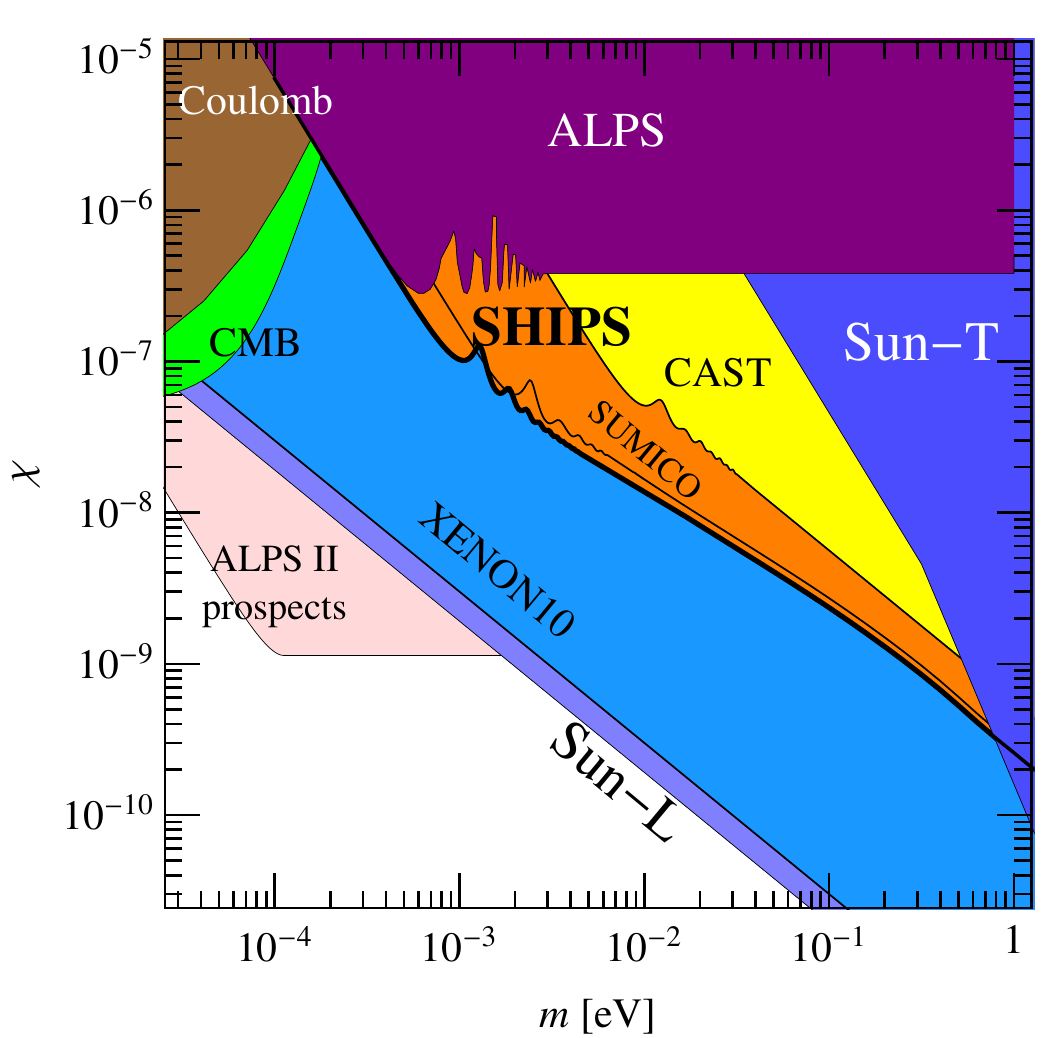}
\caption{Parameter plane of HP mass and mixing parameter. The 95\% confidence level upper limit on the mixing parameter as a function of mass estimated by the SHIPS experiment is marked orange and labeled as `SHIPS'. The coloured areas are excluded by theoretical considerations or other experiments. The latter are the Light Shining through
Walls experiment Any Light Particle Search (ALPS), the dark matter search experiment `XENON10'~\cite{An:2013yua} and FIRAS (Far Infrared Absolute Spectrophotometer) marked `CMB'. Excluding considerations from precision measurements of Coulomb's law are marked as `Coulomb' and the solar luminosity constraints in the longitudinal channel as `Sun-L'.}
\label{plane}
\end{figure}

\begin{figure}[h]
\centering
\includegraphics[width=90mm]{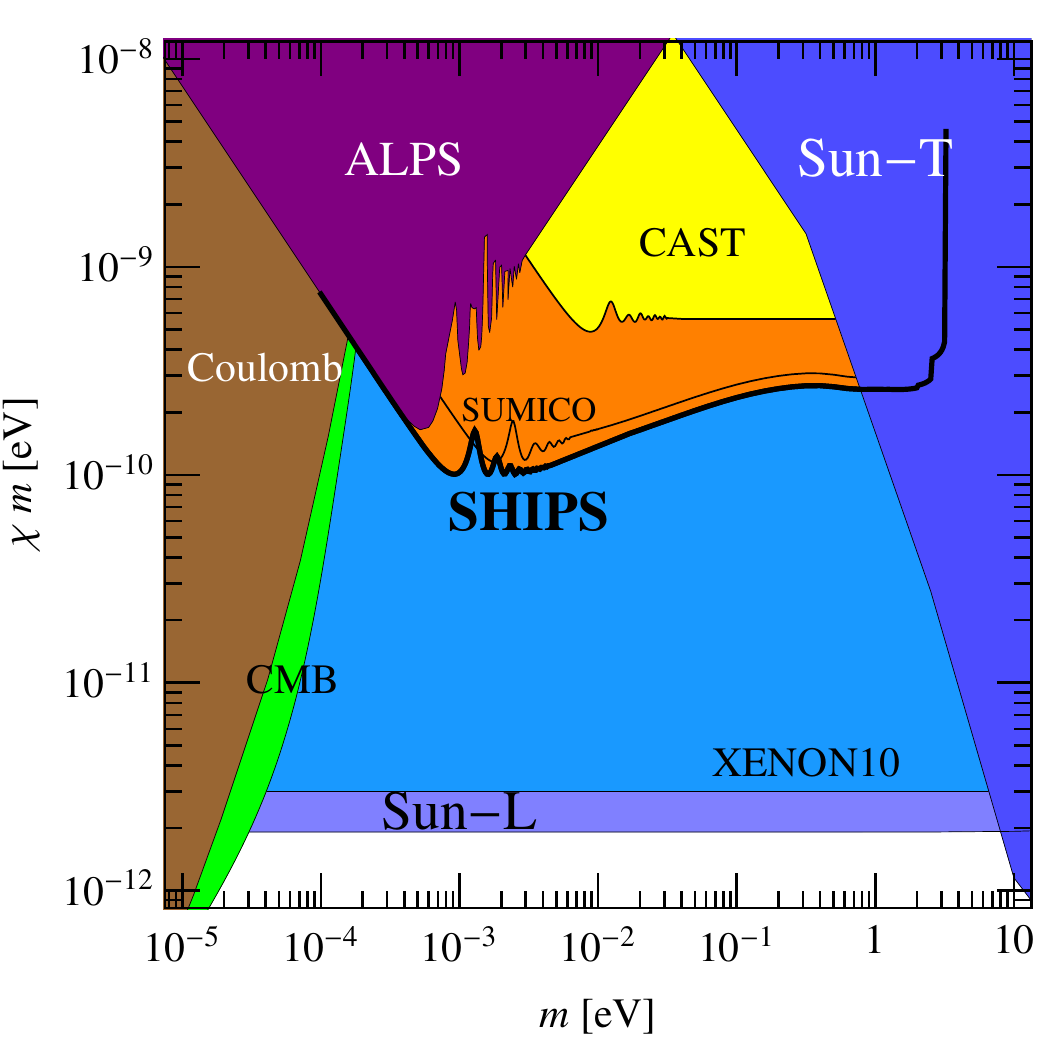}
\caption{Same as Fig.~\ref{plane} but showing the constraints on $\chi m$, which tend to level out at low masses.}
\label{plane2}
\end{figure}

\begin{figure}[t]
\centering
\includegraphics[width=90mm]{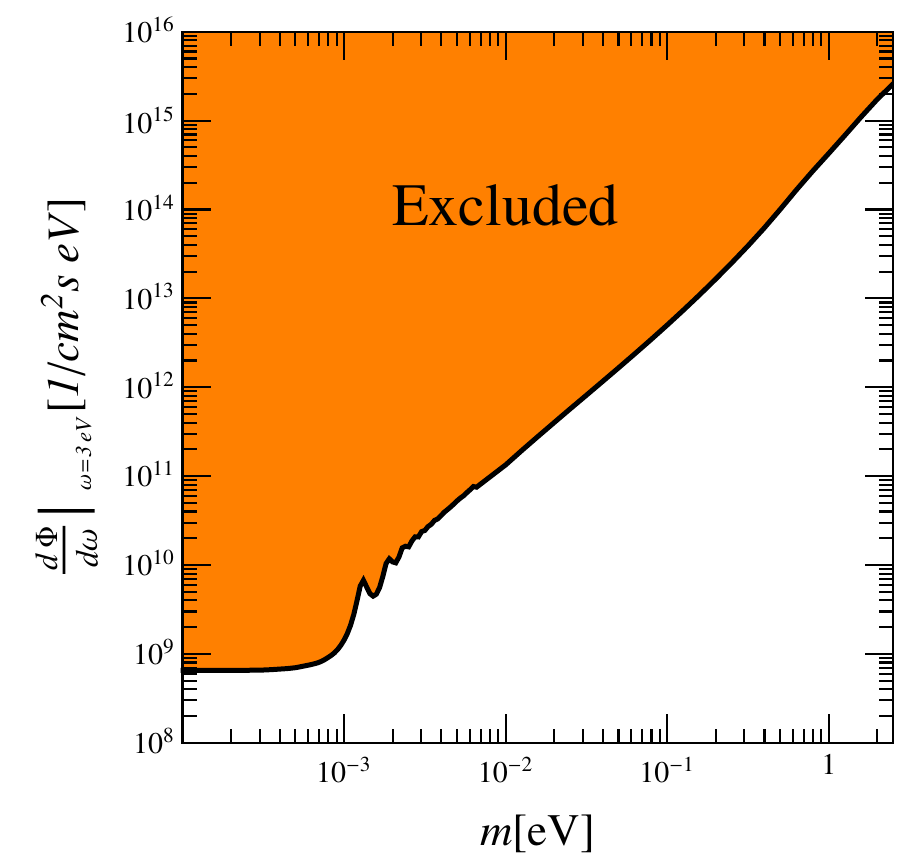}
\caption{Constraint on the flux of solar hidden photons as a function of HP mass obtained from the TSHIPS experiment data presented in this paper.} 
\label{HPfluxconstraint}
\end{figure}

In Fig.~\ref{plane}, we show the value of $\chi$, which saturates our 95\%C.L. detection rate as a function of the HP mass, as a black line labelled SHIPS. The orange region at larger values of $\chi$ is thus excluded. To ease the comparison with other experiments, we provide a completely analogous plot of $\chi m$ in Fig.~\ref{plane2}. Based on the estimation of the flux, we can also constrain the flux of HPs with $\omega \sim 3$ eV in TSHIPS, see Fig.~\ref{HPfluxconstraint}.  

The recent limits from the XENON10 analysis~\cite{An:2013yua} and solar precision tests~\cite{An:2013yfc,Redondo:2013lna,Vinyoles:2015aba}, based on the more abundant flux of longitudinally-polarised HPs, are much stronger than the recent TSHIPS limits. Note, however, that SHIPS provides the most restrictive experimental exclusion limits in the transverse-mode. In particular, our results are more sensitive than the solar-neutrino argument~\cite{Redondo:2013lna} (in T-modes) and thus are compatible with neutrino observations and self-consistent.

At this point it is interesting to speculate about the necessary modifications of our experiment that would enable us to compete with the sensitivity of XENON10 and solar precision tests. It is apparent from Fig.~\ref{plane2} that SHIPS requires a boost of $\sim$2 orders of magnitude in sensitivity to $\chi$ (at least above $m\sim$ meV). Since the signal depends on $\chi^4$, this means an improvement of 8 orders of magnitude in the signal to noise ratio.  
We think that this is only possible if a future search focuses on the infrared flux, which is larger (although more uncertain). 
The coupling of a large aperture lens to the narrow chip would be certainly a demanding challenge. Interestingly, if one focuses in the infrared, the oscillation length turns out to be uncritical, the suppression of the first oscillation would happen only for $m<\sqrt{2 \omega/\pi L}$ which is only $\sim 4\times 10^{-4}$ eV for $\omega=0.5$ eV and the length of TSHIPS.  
Assuming thus a $4.15$-m oscillation length, a stable dark-count rate of $10^{-4}$ Hz, $100\%$ detection efficiency and one year of measurement time, we find that the aperture area required to compete with the XENON10 sensitivity is around 10 m$^2$.  

\section{Conclusions and Outlook}\label{sec:conclu}

The `Solar Hidden Photon Search' was performed successfully. Our helioscope named TSHIPS is the most sensitive instrument for a HP search in the transverse-mode at present. Its set-up and measuring method provided clean and proper conditions for a sub-eV mass hidden photon detection. The SHIPS data was gained in complementary series of Sun and background measurements. The very strict equality of both enables us to constrain the flux of photons regenerated from solar HPs inside TSHIPS to be smaller than 25 mHz/m$^2$. With this figure and the most recent estimation of the solar HP flux we imposed new constraints on the HP mixing-mass plane. 

The approach to explore uncharted HP parameter space with more efficient (longer, wider, quieter) helioscopes is technically and financially no longer straightforward. Thus totally new considerations and approaches have to be excogitated to progress in the search for hidden photons and a new fundamental force of nature. 

\section*{Acknowledgements}  
The authors like to thank Rayk Nachtigall for his support concerning photomultipliers, according read-out-equipment and the provision of a custom photomultiplier test stand. We also like to thank Ulrich K\"otz and Wladimir Hain at DESY for their assistance with the troubleshooting at our photomultiplier. We acknowledge the support by the Collaborative Research Center 676 `Particles, Strings, and the Early Universe' and the Hamburg Excellence Initiative `Connecting Particles with the Cosmos'. Finally we thank sincerely Stefan Dahmke for his enduring aid in the set-up, maintenance or other concerns since the beginning of the SHIPS project. The work of J. R. was supported by the the Alexander von Humboldt Foundation, Deutsche Forschungsgemeinschaft through grant No. EXC 153, the European Union through the Initial Training Network `Invisibles' PITN-GA-2011-28944 and the Spanish nations through the Ram\'on y Cajal fellowship RYC-2012-10957.

\end{document}